\documentclass[pdflatex,sn-mathphys-num]{sn-jnl}

\usepackage{graphicx}%
\usepackage{multirow}%
\usepackage{amsmath,amssymb,amsfonts}%
\usepackage{amsthm}%
\usepackage{mathrsfs}%
\usepackage[title]{appendix}%
\usepackage{xcolor}%
\usepackage{textcomp}%
\usepackage{manyfoot}%
\usepackage{booktabs}%
\usepackage{algorithm}%
\usepackage{algorithmicx}%
\usepackage{algpseudocode}%
\usepackage{listings}%
\usepackage{caption}
\usepackage{subcaption}

\theoremstyle{thmstyleone}%
\theoremstyle{thmstyletwo}%

\theoremstyle{thmstylethree}%

\begin{document}

\title[Stability analysis of chaotic systems in latent spaces]{Stability analysis of chaotic systems in latent spaces}


\author[1]{\fnm{Elise} \sur{\"Ozalp}}\email{elise.ozalp@imperial.ac.uk}

\author*[1, 2,3]{\fnm{Luca} \sur{Magri}}\email{l.magri@imperial.ac.uk}

\affil[1]{\orgdiv{Department of Aeronautics}, \orgname{Imperial College London}, \orgaddress{\street{South Kensington Campus}, \city{London}, \postcode{SW7 2BX,}, \country{United Kingdom}}}

\affil[2]{ \orgname{The Alan Turing Institute}, \orgaddress{\street{96 Euston Road}, \city{London}, \postcode{NW1 2DB},  \country{United Kingdom}}}

\affil[3]{\orgdiv{DIMEAS}, \orgname{Politecnico di Torino}, \orgaddress{\street{Corso Duca degli Abruzzi}, \city{Torino}, \postcode{ 24 10129}, \country{Italy}}}


\abstract{
Partial differential equations, and their chaotic solutions, are pervasive in the modelling of complex systems in engineering, science, and beyond. Data-driven methods can find solutions to partial differential equations with a divide-and-conquer strategy: The solution is sought in a latent space, on which the temporal dynamics are inferred (``latent-space'' approach). This is achieved by, first, compressing the data with an autoencoder, and, second, inferring the temporal dynamics with recurrent neural networks. The overarching goal of this paper is to show that a latent-space approach can not only infer the solution of a chaotic partial differential equation, but it can also predict the stability properties of the physical system.
First, we employ the convolutional autoencoder echo state network (CAE-ESN) on the chaotic Kuramoto-Sivashinsky equation for various chaotic regimes. We show that the CAE-ESN (i) finds a low-dimensional latent-space representation of the observations and (ii) accurately infers the Lyapunov exponents and covariant Lyapunov vectors (CLVs) in this low-dimensional manifold for different attractors. Second, we extend the CAE-ESN to a turbulent flow, comparing the Lyapunov spectrum to estimates obtained from Jacobian-free methods. 
A latent-space approach based on the CAE-ESN effectively produces a latent space that preserves the key properties of the chaotic system, such as Lyapunov exponents and CLVs, thus retaining the geometric structure of the attractor. The latent-space approach based on the CAE-ESN is a reduced-order model that accurately predicts the dynamics of the chaotic system, or, alternatively, it can be used to infer stability properties of chaotic systems from data. 
}




\maketitle
\section{Introduction}
Chaotic systems arise in applications in various fields, from meteorology \citep{lorenz63} to propulsion \citep{huhn2020stability}, with turbulence frequently cited as a classic example of spatiotemporal chaotic behaviour \citep{takens1981detecting}. Data-driven reduced-order modelling has become a promising approach for learning solutions of high-dimensional chaotic partial differential equations \citep{racca2023predicting, linot2022data, nakamura2021convolutional, borrelli2022predicting, mata2023forecasting}  by leveraging the fact that, whilst the state may evolve in a high-dimensional space, the long-term dynamics often converge to a lower-dimensional attractor. A computationally efficient strategy employs a sequential approach: first, by identifying a low-dimensional manifold, or latent space, that captures the core coordinates, and second, by forecasting the temporal evolution within the latent manifold. One successful framework employing this strategy is to use a convolutional autoencoder (CAE) \citep{kramer1991nonlinear} to compress the data into a nonlinear latent representation and a recurrent neural network to propagate the lower-dimensional temporal dynamics on the latent manifold.

The CAE is a robust data-driven order reduction method that computes a manifold representation of the data and outperforms linear compression methods such as proper orthogonal decomposition \citep{berkooz1993proper}; but at the loss of interpretability of modes and coherent structures \citep{mo2024decoder}. The generated latent space represents a spatial compression of the data, and the aim is to forecast the temporal evolution within this latent manifold. Recurrent neural networks are architectures used to predict dynamics based on a time series input: they have been successfully applied to full-state chaotic systems \citep{Margazoglou2023stability, Vlachas_2020_backprop, Pathak_2017_ml_le, Pathak_2018_modelfree_pred}, to name a few. As hybrid architectures, recurrent neural networks have been coupled with AE to model chaotic systems \citep{vlachas2022multiscale, gupta2023mori} and fluid flows \citep{racca2023predicting, hasegawa2020cnn}. Although these methods have demonstrated particular success in short-term forecasting, their long-term properties and the interpretability of the latent manifolds generated by the CAE remain largely unexplored \citep{magri2022interpretability}.

To develop a physically accurate surrogate model for chaotic dynamics, it is essential to capture the properties of the chaotic attractor \citep{ozalp2023reconstruction}. These attractors are characterized by invariant measures such as Lyapunov exponents, the Kaplan-Yorke dimension, and the geometry of the tangent space. The predictability and stability of a chaotic system are intrinsically linked to its tangent space, which can be analysed through the linearized dynamics provided by the Jacobian. However, in large turbulent systems, the high-dimensional Jacobian becomes the bottleneck in the calculation of Lyapunov exponents. This has led to the adoption of Jacobian-free methods for estimating separation growth rates, which, whilst computationally cheaper, still require knowledge of the governing equations and a suitable numerical solver \citep{dieci2002jacobian}. Although stability properties can be inferred using techniques like echo state networks (ESNs) \citep{Margazoglou2023stability} and long short-term memory networks \citep{ozalp2023reconstruction}, these methods are not well-suited for high-dimensional data. This results in two key challenges: (i) the stability properties of data-driven reduced-order models remain poorly understood, and (ii) existing stability analysis methods are inadequate for high-dimensional data.

In this paper, we address this gap by employing ESNs, a reservoir computer, as recurrent neural networks \citep{jaeger2001echo} and applying stability analysis to investigate and characterize latent manifolds, thus extending the CAE-ESN methodology \citep{racca2023predicting}. First, we apply the CAE-ESN methodology to Kuramoto-Sivashinsky equation and analyse the stability of the latent dynamics using the ESN Jacobian. We compute the Lyapunov exponents, the Kaplan-Yorke dimension, and the angles of the covariant Lyapunov vectors in the latent manifold, and compare this to the numerical reference based on the Jacobian of the equation. Second, we assess the robustness of the CAE-ESN stability analysis across varying parameters and dimensions of attractors. Third, we extend the CAE-ESN methodology to the quasiperiodic and turbulent Kolmogorov flow, calculating Lyapunov exponents and comparing this approach to a Jacobian-free method. 

This paper is organised as follows. Key concepts essential for stability analysis for chaotic and turbulent systems are introduced in Section~\ref{sec:stability_properties}. Section~\ref{sec:cae-esn} elaborates on the CAE-ESN methodology. In Section~\ref{sec:ks-results}, the CAE-ESN is applied to the Kuramoto-Sivashinsky equation, demonstrating the full stability analysis on various parameter configurations. The CAE-ESN is extended to the quasiperiodic and turbulent Kolmogorov flow in Section~\ref{sec:kolmogorov-results}. Finally, Section~\ref{sec:conclusion} summarises our findings.

\section{Stability Properties}\label{sec:stability_properties}
In this section, we introduce the four key concepts from the stability of chaotic systems, which are used in this paper: the Lyapunov spectrum, the Kaplan-Yorke dimension, covariant Lyapunov vectors, and inertial manifolds. For more details, the reader is referred to \citep{strogatz2018nonlinear, Sandri_1996_NumericalCalculationLyapunovExponents, ruelle1979ergodic, kuptsov2012theory, Magri2024vki}. 
We consider an autonomous dynamical system 
\begin{align}\label{eq:dyn_sys}
\mathbf{\Dot{u}}(t) = f(\mathbf{u}(t))
\end{align}
with $\mathbf{u}(t)  \in \mathbb{R}^{N}$ and $f \in \mathbf{C}^1(\mathbb{R}^{N})$. When considering partial differential equations (PDEs), and in this paper dissipative PDEs, $f$ can be treated as the spatially discretized operator with embedded boundary conditions. Given an initial condition $\mathbf{u}_0 = \mathbf{u}(t_0)$, the solution of Eq.~\eqref{eq:dyn_sys} is the trajectory $\mathbf{u}(t)$  in the phase space of dimension $N$.
When the system in Eq.~\eqref{eq:dyn_sys} is perturbed by an infinitesimal perturbation $\boldsymbol{\varepsilon}_0 $ on the initial condition, i.e. $\mathbf{u}_0 + \boldsymbol{\varepsilon}_0$, the  temporal evolution of the perturbation is  
\begin{equation}\label{eq:tangent_equation}
    \frac{d\boldsymbol{\varepsilon}(t)}{dt} = \mathbf{J}(\mathbf{u}(t))\boldsymbol{\varepsilon}(t),
\end{equation}
 where $J_{ij} = \partial f_i(\mathbf{u}) / \partial u_j$ are the components of the Jacobian $\mathbf{J}(\mathbf{u}(t)) \in \mathbb{R}^{N\times N}$. Eq.~\eqref{eq:tangent_equation} shows that the perturbation $\mathbf{\epsilon}(t)$ evolves along the (time-varying) tangent space centred at $\mathbf{u}(t)$.
 
Chaotic systems are characterized by the exponential divergence of infinitesimally close initial conditions. Specifically, any perturbation norm $\| \boldsymbol{\varepsilon}(t)\|$ grows exponentially for $t \to \infty$, following $\| \boldsymbol{\varepsilon}(t)\| \sim e^{\lambda_1 t} \| \boldsymbol{\varepsilon}_0\|$, where $\lambda_1$ is the leading Lyapunov exponent  (LE) \citep{strogatz2018nonlinear, Sandri_1996_NumericalCalculationLyapunovExponents}. The leading Lyapunov exponent characterizes the dynamics of the system. If $\lambda_1<0$, the perturbation decays and the attractor is a fixed point. On the other hand, the attractor is a quasiperiodic torus of dimension $j$ if $\lambda_i=0$ for $i \leq j$ and $\lambda_i < 0 $ for $i>j$.  If $\lambda_1> 0 $, the perturbation grows exponentially and, typically, the attractor is chaotic. 
In chaotic systems, the inverse of the leading Lyapunov exponent, $\tau_{\lambda} = \frac{1}{\lambda_1}$, provides a characteristic timescale for two nearby trajectories to separate. This timescale $\tau_{\lambda}$ is used to assess the system’s predictability horizon and is referred to as the Lyapunov time. The leading Lyapunov exponent is the dominant exponent of the Lyapunov spectrum. From the multiplicative ergodic theorem, it can be shown that there exist  $\lambda_1 \geq \dots \geq \lambda_{N}$ Lyapunov exponents in the limit of $t \to \infty$ \citep{Oseledec_1968_MultiplicativeErgodicTheorem}, the collection of which is the Lyapunov spectrum. Physically, the Lyapunov exponents are the average exponential contraction/expansion rates in the $N$ directions of the phase space along the attractor. Finally, the knowledge of the Lyapunov exponents offers an upper-bound estimate of the attractor's dimension, defined by the Kaplan-Yorke dimension \citep{frederickson1983liapunov}
\begin{equation}
    D_{KY} = k + \frac{\sum_{i=1}^k \lambda_i}{|\lambda_{k+1}|}
\end{equation}
 with $\sum_{i=1}^k \lambda_i > 0 $ and $\sum_{i=1}^{k+1} \lambda_i < 0$. The attractor's dimension is a fundamental characteristic of chaotic systems, which represents the minimum number of degrees of freedom required to model the system's dynamics accurately. 
 
 \subsection{Geometry of the tangent space}
To gain insight into the geometry of the attractor, covariant Lyapunov vectors (CLVs) are useful invariant quantities. CLVs are associated with the Lyapunov exponents and are covariant with the tangent flow, i.e. the $i$-th CLV at time $t_1$ is mapped to the $i$-th CLV at $t_2$. The CLVs $\mathbf{V}= \begin{bmatrix} \mathbf{v}_1, \mathbf{v}_2, \dots, \mathbf{v}_N\end{bmatrix}$ are uniquely defined (i.e. non-degenerate), and provide a (generally) non-orthogonal, local splitting of the phase space into three subspaces. The unstable, expanding subspace $E^U_{\mathbf{u}}$ is associated with the positive Lyapunov exponents, the neutral subspace $E^N_{\mathbf{u}}$ describes the space spanned by the CLVs associated with zero Lyapunov exponents and the stable and contracting subspace $E^S_{\mathbf{u}}$ corresponds to the negative Lyapunov exponents ~\citep{Oseledec_1968_MultiplicativeErgodicTheorem, ruelle1979ergodic}. 
When these subspaces are strictly separated, i.e. there are no tangencies, the system is hyperbolic. This implies structural stability and linear response in hyperbolic chaos that persists under changes in parameters, which is relevant for shadowing methods and adjoint sensitivity analysis \citep{chater2017least, palis1995hyperbolicity, huhn2020stability}.
To calculate the splitting, the angle between vectors from these subspaces for different points along trajectories must be bounded away from zero. This can be achieved by computing the principal angles of the intersection between  $E^U_{\mathbf{u}}$, neutral $E^N_{\mathbf{u}}$, and stable $E^S_{\mathbf{u}}$, which is a resource-intensive process. However, even calculating the distributions of the angle $\theta$ between pairs of CLVs can provide insight into the geometry of the attractor. The angles between a pair of CLV $\mathbf{v}_i, \mathbf{v}_j$ are computed as 
 \begin{equation}\label{eq:clv_angles}
	\theta_{\mathbf{v}_i, \mathbf{v}_j} = \frac{180^{\circ}}{\pi} \cos^{-1}(|\mathbf{v}_i \cdot  \mathbf{v}_j|)
\end{equation} 
where $\cdot$ denotes the dot product \citep{Ginelli2007_CharacterizingDynCLVs}. 
%
\subsection{Inertial manifold of dissipative systems}
For dissipative PDEs, the long-term behaviour of the system is described by the inertial manifold, which is an exponentially attracting smooth manifold embedded in the attractor. The inertial manifold provides a finite-dimensional representation of the fractal attractor, providing a more practical framework for reduced-order modelling \citep{constantin2012integral, foias1988inertial}. To numerically accurately integrate the PDE, the number of degrees of freedom for the simulation is given by the inertial manifold's dimension, which can be up to twice the Kaplan-Yorke dimension \citep{yang2009hyperbolicity}.
Through the inertial manifold, the tangent space is split into two decoupled parts: a set of strongly interacting physical modes spanning the manifold, and a separate set of spurious modes that are hyperbolically decoupled. The number of physical modes remains constant in the high resolution limit of the PDE, whilst the spurious modes typically arise from increasing spatial resolution and are associated with very negative Lyapunov exponents. The inertial manifold reduces the system's dynamics by filtering out spurious modes and retaining only the physical modes such that the  key dynamics necessary for accurate modelling and prediction are retained.

This splitting can be identified by computing the distribution of CLV pairs \citep{takeuchi2011hyperbolic, yang2009hyperbolicity}. First, the CLVs are collected over a sufficient time to ensure convergence. Second, the angle distribution between consecutive pairs is calculated, as described in Eq.~\eqref{eq:clv_angles}, to examine the absence of tangencies. These indicate the hyperbolic decoupling of the inertial manifold from the spurious space. Lyapunov exponents alone are insufficient to identify this hyperbolic decoupling; instead, the key quantities required are the CLVs.

\subsection{Stability analysis of turbulent systems}

Turbulence is often viewed as a typical example of spatiotemporal chaos with large numbers of degrees of freedom. In this section, we discuss the adaptation of stability analysis methods, specifically Jacobian-free approaches, to address the challenges posed by the high dimensionality of turbulent systems.

Solving the Navier-Stokes equations for high Reynolds numbers requires small time steps and fine spatial resolution for numerical stability. However, the large number of degrees of freedom becomes the bottleneck to approximating Lyapunov exponents because one has to evaluate the Jacobian and subsequently propagate the tangent equation through matrix-vector multiplication. This has led to the use of Jacobian-free methods for estimating Lyapunov exponents as separation growth rates \citep{bec2006lyapunov,labahn2019determining, nastac2017lyapunov}. These methods, whilst computationally cheaper, require knowledge of the governing equations with the corresponding numerical solver. 

To estimate $m$ Lyapunov exponents, the system's trajectory is perturbed by $m$ perturbations, which are evolved over short time intervals. By tracking the growth of these perturbations and periodically reorthonormalizing them, the growth rates can be approximated, providing an estimate of the Lyapunov exponents. For further details, we refer to \citep{Sandri_1996_NumericalCalculationLyapunovExponents, dieci2002jacobian, racca2023predicting}. The iterative procedure is repeated until the spectrum is converged. This method can be time-intensive and typically results in an approximation of the Lyapunov spectrum, which is prone to inaccuracies in the negative Lyapunov exponents \citep{pikovsky2016lyapunov}. Furthermore, the difficulty in calculating Lyapunov exponents extends to the computation of CLVs as well. For the specific system studied in this work —the Kolmogorov flow—previous research has explored its hyperbolicity and performed covariant Lyapunov analysis \citep{inubushi2012covariant}. It has been found that the system is hyperbolic at low Reynolds numbers and transitions to a non-hyperbolic state at certain higher Reynolds numbers. The Navier-Stokes attractor reflects the nonlinearity of turbulence from the large to the small spatial structures of the flow. Although the existence of an exact inertial manifold is not generally proven, approximate inertial manifolds for 2D and 3D turbulence exist \citep{temam1989inertial, foias1991approximate} -  providing a motivation to seek a lower dimensional representation which captures the essential turbulent dynamics.

\section{Learning the dynamics of the inertial manifold}\label{sec:cae-esn}
\begin{figure}[h]
    \centering
    \includegraphics[width=\linewidth]{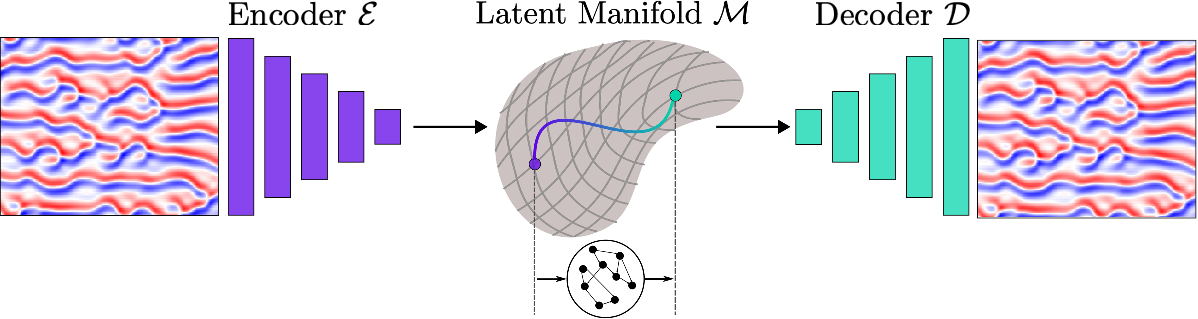}
    \caption{Illustration of the CAE-ESN applied to the Kuramoto-Sivashinsky equation. The encoder maps the physical state to a low-dimensional latent manifold, where the ESN propagates the dynamics. The decoder then decompresses the predicted latent representation to the respective high-dimensional physical state.}
    \label{fig:cae_illustration}
\end{figure}

For dissipative chaotic systems as described in Eq.~\eqref{eq:dyn_sys}, although the state has many degrees of freedom; in the asymptotic limit, the solution converges to a chaotic attractor with a smaller dimension. This attractor is a subset within the $N$-dimensional phase space, characterized by a lower-dimensional fractal dimension. The existence of this lower-dimensional manifold $\mathcal{M}$, provides a principled justification to reduce the complexity of the model without affecting the accuracy. In other words, the ansatz of this work is to compute the low-dimensional manifold using a convolutional autoencoder (CAE), which offers a nonlinear reduced-order representation of the solution (Fig.~\ref{fig:cae_illustration}). 
By mapping each snapshot of the physical space onto the latent space, we obtain a low-dimensional time series, whose temporal dynamics can, in turn, be inferred by another network. Following this approach, two networks are trained for the spatial and temporal dynamics separately \citep{racca2023predicting}. Whilst it is possible to train these networks in an end-to-end approach \citep{vlachas2022multiscale}, the computational cost for training is significantly lowered by first training the autoencoder and then training the model for the temporal dynamics. Although previous work employs long short-term memory networks \citep{vlachas2022multiscale}, neural ordinary differential equations \citep{linot2022data} and conditional random fields \citep{herzog2019convolutional} in the latent space, we follow the approach of \citep{racca2023predicting} by using the echo state network (ESN) as a recurrent neural network. This is because, as opposed to gated architectures, ESNs are trained by solving a simple quadratic optimization problem, which does not require backpropagation. Further, their recurrent structure allows for straightforward data-driven stability analysis \citep{Pathak_2017_ml_le, Margazoglou2023stability}. 

\subsection{Convolutional autoencoders}\label{subsec:cae}
Autoencoders \citep{hinton1993autoencoders, vincent2008extracting} are neural networks designed to approximate the identity operator, enabling efficient encoding tasks such as feature extraction and dimensionality reduction \citep{wang2014generalized} by effectively compressing observations into lower-dimensional manifolds \citep{fefferman2016testing}.

%
 An autoencoder consists of an encoder and a decoder. The encoder, denoted by $\mathcal{E}$, maps the physical input $\mathbf{u}(t_i) \in \mathbb{R}^{N}$ to a latent representation $ \mathbf{y}(t_i) \in \mathbb{R}^{N_{lat}}$ (on the manifold $\mathcal{M}$). The decoder $\mathcal{D}$ reconstructs the physical state from the latent representation by mapping back to the full state such that
\begin{align}
    \mathbf{\hat{u}}(t_i) \approx \mathbf{u}(t_i), \quad \text{where } \mathbf{\hat{u}}(t_i) = \mathcal{D}\left( \mathbf{y}(t_i) \right), \quad \mathbf{y}(t_i)=\mathcal{E}\left(\mathbf{u}(t_i) \right),
\end{align}
and $\mathbf{\hat{u}}(t_i)$ is the autoencoder reconstruction. In this paper, the network is trained by minimizing the mean squared reconstruction error (MSE)
\begin{align}
    \mathcal{L}(\mathbf{u}, \mathbf{\hat{u}}) = \frac{1}{N_{train}} \sum_{i=1}^{N_{train}} \| \mathbf{u}(t_i) - \mathbf{\hat{u}}(t_i) \|_2^2, 
\end{align}
where $N_{train}$ denotes the number of training samples. By employing convolutional layers \citep{lecun1989generalization} in the encoder and decoder, the spatial structure of the input is taken into account, resulting in a convolutional autoencoder (CAE). Nonlinear activation functions follow each convolutional layer; linear activation functions would render the network equivalent to principal component analysis (PCA) \citep{mo2024decoder}. However, nonlinear compression is essential for extracting a minimal representation of the nonlinear manifold.
Generally, the reconstruction error on the test data decreases as the latent dimension increases until it eventually converges for larger dimensions. However, determining an appropriate latent space dimension is not straightforward and depends on the network architecture. The dimension of the chaotic attractor serves as a strict lower bound for the latent space dimension\citep{racca2023predicting}, but is typically unknown. Estimation of the attractor's dimension can be achieved through the Kaplan-Yorke dimension \citep{farmer1983attractordimension}. Although this estimation provides a lower bound when the equations are known, the latent dimension often needs to be selected larger, particularly because the encoder introduces approximation errors.
In spatially dissipative systems, the attractor is embedded within the inertial manifold, where the number of modes spanning this manifold can extend to approximately twice the Kaplan-Yorke dimension \citep{yang2009hyperbolicity} 
In this paper, we argue that the physical modes encapsulate the essential information regarding phase dynamics, which should be captured by an optimal low-dimensional latent representation. Related works find a significant drop in the reconstruction of the test error at the specific latent dimension corresponding to the number of physical modes \citep{linot2022data, vlachas2022multiscale}.

\subsection{Echo state networks}\label{subsec:esn}
The trained autoencoder provides a spatial transformation from the physical snapshot $\mathbf{u}(t_i)$ to the latent representation $\mathbf{z}(t_i)$. 
For learning the temporal dynamics of the training data, we employ the echo state network (ESN)\citep{jaeger2001short} on the time-ordered latent representations $\{ \mathbf{y}(t_i)\}_{i=1, \cdots, N_{train}}$, and the ESN implementation follows \citep{lukovsevivcius2012practical, racca2021robust, Margazoglou2023stability, racca2023predicting}. At each time step, the ESN maps the latent state input $\mathbf{y}(t_i)$ to the reservoir state
\begin{align}\label{eq:esn1}
        \mathbf{r}(t_{i+1}) &= \tanh\left([\mathbf{y}(t_i); b_{in}]^T\mathbf{W}_{in} + \mathbf{r}(t_i)^T\mathbf{W} \right).
\end{align}
The reservoir is employed to compute the predicted latent state at the next time step $  \mathbf{\Hat{y}}(t_{i+1})$, 
\begin{align}\label{eq:esn2}
 \mathbf{\Hat{y}}(t_{i+1}) & = [\mathbf{r}(t_{i+1}), b_{out}]^T \mathbf{W}_{out}.
\end{align}
\begin{figure}[h]
    \centering
    \includegraphics[width=0.5\textwidth]{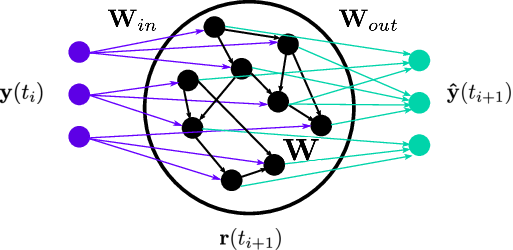}
    \caption{Schematic representation of an echo state network\citep{Margazoglou2023stability}.}
    \label{fig:esn_cell}
\end{figure}

\begin{figure}[h]
     \centering
     \begin{subfigure}[b]{0.47\textwidth}
         \centering
        \includegraphics[width=\textwidth]{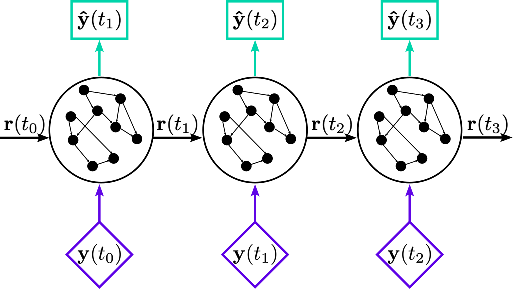}
        \caption{Open-loop configuration of the echo state  network\citep{Margazoglou2023stability}.}
    \label{fig:esn_open_loop}
     \end{subfigure}
     \hfill
     \begin{subfigure}[b]{0.47\textwidth}
        \centering
        \includegraphics[width=\textwidth]{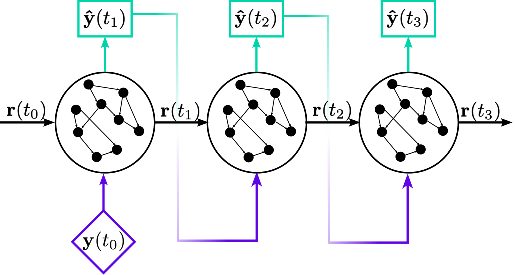}
        \caption{Closed-loop configuration of the echo state network\citep{Margazoglou2023stability}.}
        \label{fig:esn_closed_loop}
     \end{subfigure}
\end{figure}

The matrices $\mathbf{W}_{in}$ and $\mathbf{W}$ are (pseudo-) randomly generated and fixed, whilst the weights of the output matrix, $\mathbf{W}_{out}$, are computed during training. The input matrix $\mathbf{W}_{in} \in \mathbb{R}^{N_r \times (N_{lat} + 1)}$ is dense and the entries are sampled from a uniform distribution ${U}[-\sigma_{in}, \sigma_{in}]$, where $\sigma_{in} \in \mathbb{R}$ is the input scaling. The state matrix $\mathbf{W} \in \mathbb{R}^{N_r \times N_r}$ is an Erd\"os-Renyi matrix with average connectivity $d$. This means that on average, each neuron (each row) of $\mathbf{W}$ has only $d$ connections (non-zero) elements, resulting in a sparse matrix. The connections are sampled uniformly from $[-1, 1]$, then rescaled to a set spectral radius $\rho$ to fulfil the echo state property \citep{jaeger2001echo}. The biases $b_{in}, b_{out}$ are added to break the symmetry of the network \citep{Herteux_2020_breaksymRC}.
The ESN can be employed in two modes, open-loop and closed-loop. During training, the network operates in open-loop mode, during which it predicts the next time step based on a reference input (Fig.~\ref{fig:esn_open_loop}), after discarding an initial transient (the washout interval) \citep{racca2021robust}. Using the open-loop configuration, which follows Eq.~\ref{eq:esn1}, the output matrix $\mathbf{W}_{out} \in \mathbb{R}^{(N_r+1) \times N_{lat}}$ is trained by minimizing the mean squared error (MSE) between the network prediction $ \mathbf{\Hat{y}}(t_{i})$ and the reference data $\mathbf{y}(t_i)$. It can be shown that this is equivalent to solving for $\mathbf{W}_{out}$ via ridge regression of 
\begin{align}
    (\mathbf{R}\mathbf{R}^T + \beta \mathbb{I}) \mathbf{W}_{out} = \mathbf{R}\mathbf{Y}^T,
\end{align}
where $\mathbf{R}\in \mathbb{R}^{(N_r+1) \times N_{train}}$ and $\mathbf{Y}^T$ are the horizontal concatenations of the reservoir states (with bias) and the output data, respectively, $\mathbb{I}$ is the identity matrix, and $\beta$ is the Tikhonov regularisation parameter \citep{tikhonov1995numerical}, which is selected during hyperparameter optimization. The hyperparameters are selected during validation and follow \citep{racca2021robust}. 

Once $\mathbf{W}_{out}$ has been fixed, the network can be employed in the closed-loop configuration (Fig.~\ref{fig:esn_closed_loop}). In closed-loop mode, the network prediction is used as input in Eq.\eqref{eq:esn1}, allowing the network to autonomously evolve without additional input data. This setup effectively defines a dynamical system, and stability properties of the network can be calculated\citep{Margazoglou2023stability}. The first step is to calculate the Jacobian of the ESN, given by

\begin{equation}\label{eq:esn_jac}
    \mathbf{J}_{esn}(\mathbf{r}(t_{i+1)}) = (1-\mathbf{r}(t_i)^2) \mathbf{W}_{const},
\end{equation}
where $\mathbf{W}_{const} = \mathbf{W_{in}^T \mathbf{W}_{out}^T} + \mathbf{W}^T$ is constant after training. Second, the tangent equation \eqref{eq:tangent_equation} is solved using $\mathbf{J}_{esn}$. Third, we calculate the Lyapunov exponents and CLVs from the solution of the tangent equation, following \cite{Margazoglou2023stability}. The ESN predicts the temporal dynamics on the latent manifold and the Jacobian, therefore, defines the tangent space of the latent manifold. Consequently, the calculated stability properties are inferred on the latent manifold, not in the full physical space.


\section{Kuramoto-Sivashinsky Equation}\label{sec:ks-results}

The Kuramoto–Sivashinsky (KS) equation is a fourth-order partial differential equation that models instabilities of flame fronts and represents a fundamental model in the study of spatiotemporally chaotic behaviour \citep{Kuramoto_1978_DiffusionInducedChaos, Sivashinsky_1977_NonlinearAnalysisHydrodynamicInstability}. Writing $\mathbf{u} = \mathbf{u}(t,x)$, the equation is given by
\begin{equation}\label{eq:ks_equation}
\mathbf{u}_t+ \mathbf{u}_{\mathbf{x}\mathbf{x}}+ \mathbf{u}_{\mathbf{x}\mathbf{x}\mathbf{x}\mathbf{x}}+ \mathbf{u}\mathbf{u}_{\mathbf{x}}  = 0,
\end{equation} 
with periodic boundary conditions, $\mathbf{u} (t, 0) = \mathbf{u} (t, L)$ on the spatial domain $[0, L]$. The length of domain $L$ is the bifurcation parameter and we refer to \citep{papageorgiou1991route, Hyman_1986_KuramotoSivashinsky} regarding the stability behaviour.

For the domain length $L = 2\pi\cdot 10$ as in \citep{ozalp2023reconstruction}, the system is spatio-temporally chaotic with $\lambda_1 = 0.08$. Eq.~\eqref{eq:ks_equation} is discretized with $128$ degrees of freedom and solved with a fourth-order spectral scheme for stiff PDEs \citep{Kassam_2005_fourth_order}. Starting from a random initial condition, the solution is calculated until $T = 2.5 \times 10^4$ with a time step of $\Delta t = 0.25$. After discarding an initial transient time, $T_{trans}=25$, we split the remaining data into training, validation and test set with sizes $5\times 10^4$, $2\times 10^4$ and $5\times 10^4$, respectively.
Following \citep{Margazoglou2023stability}, we calculate the tangent properties, specifically the Lyapunov exponents and CLVs. Based on the calculated Lyapunov exponents, the estimated Kaplan-Yorke dimension is $D_{KY} = 15.02$. 


\subsection{Dimensionality reduction and time series forecasting}
\begin{figure}[h]
    \centering
    \includegraphics[width=0.7\textwidth]{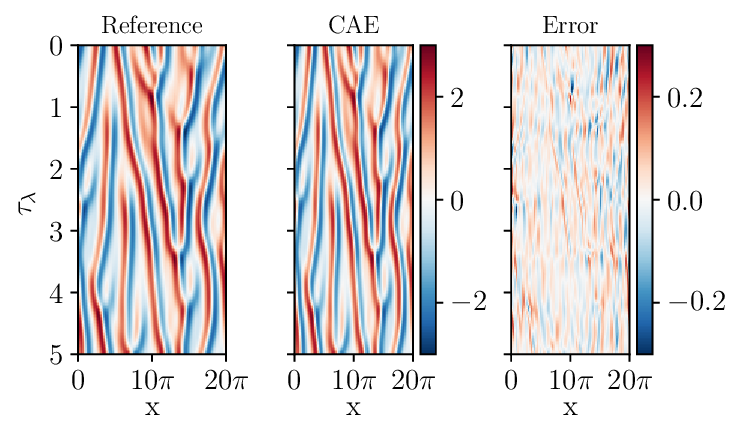} 
    \caption{Reconstruction of the Kuramoto-Sivashinsky test data with the CAE at $N_{lat}=24$.}
    \label{fig:ks_cae_24}
\end{figure}

To achieve spatial dimensionality reduction, we train dense and convolutional autoencoders with different latent dimensions (Fig.~\ref{fig:ks_cae_l60_mse}). Details of the hyperparameters can be found in Appendix \ref{sec:architecture_hyperparameter}. As a comparison, we also employ principal component analysis (PCA), a linear dimensional reduction technique that projects the training data onto an orthogonal basis of dimension $N_{lat}$. Figure.~\ref{fig:ks_cae_l60_mse} shows the reconstruction MSE on the test data of the AEs and the PCA for $N_{lat} \in [6, 8, 12, 16, 20, 24, 28, 32]$. Although the one-layer dense AE and linear PCA exhibit comparable performance on the test data, the CAE consistently outperforms both models by up to an order of magnitude. In Fig.~\ref{fig:ks_cae_24}, we present the test data and the CAE reconstruction for $N_{lat}=24$, corresponding to $~18\%$ of the degrees of freedom of the solution. 
\begin{figure}[h]
    \centering
    \includegraphics[width=0.7\textwidth]{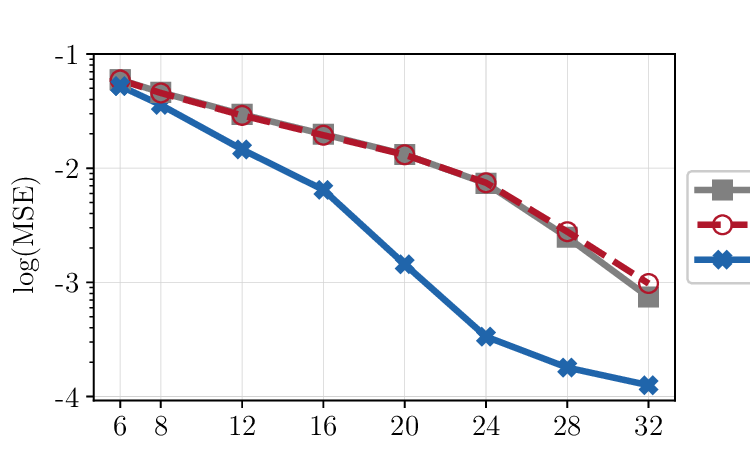} 
    \caption{Comparison of the mean squared error (MSE) in the test data reconstruction of the Kuramoto-Sivashinsky equation, across CAE, PCA, and dense AE models, as a function of increasing latent dimension.}
    \label{fig:ks_cae_l60_mse}
\end{figure}

\begin{figure}
    \centering
    \includegraphics[width=0.7\textwidth]{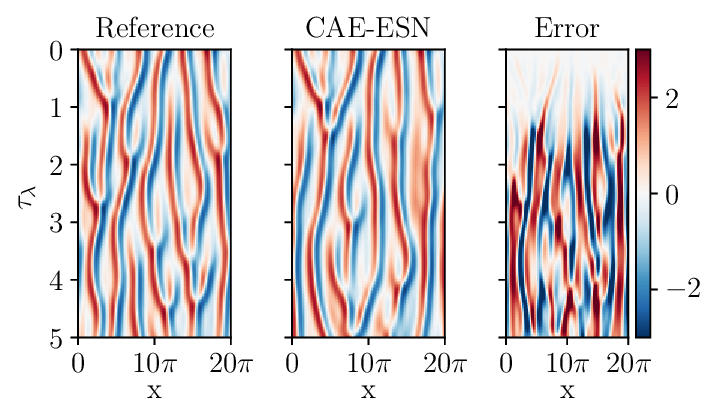} 
    \caption{Prediction of the test data of the Kuramoto-Sivashinsky equation with an ESN of $N_r=5000$ on the latent space of $N_{lat}=24$.}
    \label{fig:ks_cae_esn_24}
\end{figure}

Once the autoencoder is trained to provide the latent space, the ESN is trained on the latent space of the CAE. The closed-loop prediction of the ESN over 5 $\tau_{\lambda}$ is presented in Fig.~\ref{fig:ks_cae_esn_24}. The network accurately predicts the KS solution over a short time span, with a prediction horizon of $1.25 \tau_{\lambda}$, after which the CAE-ESN prediction diverges. Compared to previous studies \citep{Vlachas_2020_backprop}, where ESNs forecast directly on the full, physical space, the CAE-ESN forecast is based on $~18\%$ of the degrees of freedom of the solution, which is reflected in the shorter prediction horizon on the test data.

\subsection{Stability properties}\label{subsection:ks-stability}
\begin{figure}[h]
    \centering
    \includegraphics[width=0.75\textwidth]{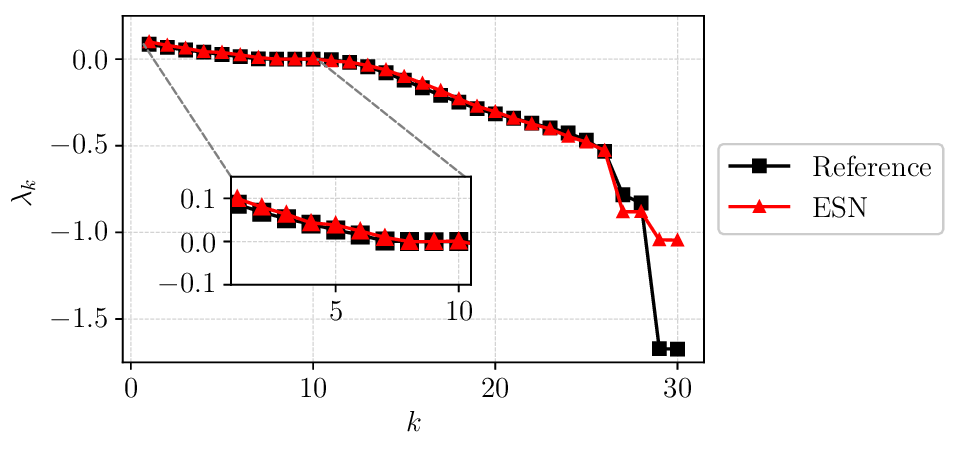} 
    \caption{Comparison between the Lyapunov exponents of the Kuramoto-Sivashinsky equation between the reference and those estimated from the ESN in the latent space.}
    \label{fig:ks_l60_cae_esn_lyap}
\end{figure}

Short-term prediction alone does not adequately gauge the physical accuracy and ability to capture the correct dynamics of the CAE-ESN. To understand the chaotic and tangent dynamics of the CAE-ESN, we perform the stability analysis outlined in Section~\ref{sec:stability_properties} after the ESN is trained. Crucially, the ESN is trained on the latent space with $N_{lat}=24$ only, i.e., it has not been trained with the full solution. The Lyapunov exponents are shown in Fig.~\ref{fig:ks_l60_cae_esn_lyap}, compared to the Lyapunov exponents calculated from the Jacobian of Eq.~\eqref{eq:dyn_sys}\footnote{The ESN does not capture the two zero Lyapunov exponents ($\lambda{9}, \lambda{10}$) as found in \citep{Vlachas_2020_backprop, ozalp2023reconstruction} and the following Lyapunov exponents are augmented accordingly.}. Overall, the mean absolute error of the first $28$ exponents is $1.5\%$, which is negligibly small. When calculating the Kaplan-Yorke dimension of the ESN, the estimated dimension is $D_{KY}^{ESN} = 15.022$ whilst the calculated reference dimension is $D_{KY} = 15.018$. This observation confirms that the latent manifold encompasses the fundamental ergodic properties of the chaotic attractor.

\begin{figure}
    \centering
    \includegraphics[width=0.7\textwidth]{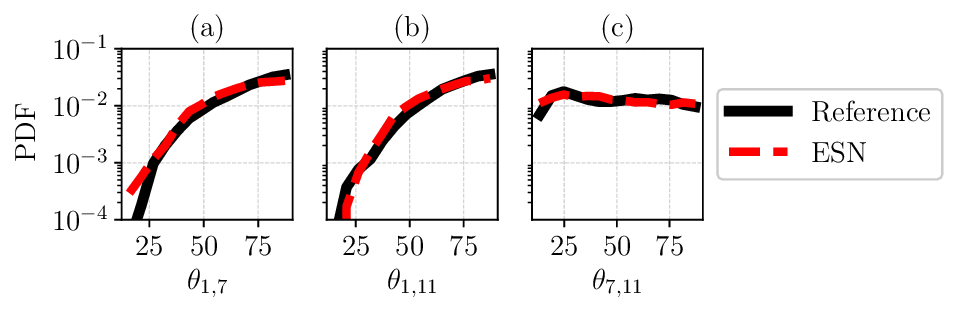} %
    \caption{The angle distribution of the Kuramoto–Sivashinsky system for the leading CLVs of three different subspaces: (a) unstable–neutral, (b) unstable–stable, and (c) neutral–stable. }
    \label{fig:ks_l60_cae_esn_clv}
\end{figure}

The stability analysis is extended to the statistics of the angles between the dominant CLVs of the unstable, neutral and stable subspace. To capture the distribution of the CLV angles, the tangent equation has to evolve significantly longer for both the reference and the ESN. In Fig.~\ref{fig:ks_l60_cae_esn_clv}, we present the CLV angles of the leading CLV calculated after Eq.~\eqref{eq:clv_angles} over $1000 \tau_{\lambda}$. The mean Wasserstein distance is $0.001$ and the agreement of angle statistics indicates that the CAE-ESN captured geometric information of the attractor in the latent space.

\subsection{Parameter study}
The CAE-ESN study is extended to different lengths of the domain of Eq.~\eqref{eq:ks_equation}. The length $L$ controls the stability of the system \citep{papageorgiou1991route} and we consider the additional cases $L=22, 44, 76$ that are chaotic. For each $L$, the CAE is trained with varying latent dimensions, followed by an ESN on the latent manifold. In all cases, we recover the Lyapunov exponents and Kaplan Yorke dimension and CLV angles by performing stability analysis on the ESN in the latent space. We present an extended analysis in the Appendix \ref{sec:appendix}. In Table~\ref{table:var_L_KS}, we summarise the key quantities, the dominant Lyapunov exponent and the Kaplan-Yorke dimension, of the reference solution and the ESN in the latent space. Across the different parameter values, the CAE-ESN require independent hyperparameter tuning. In the presented cases, the latent spaces of CAE-ESN successfully capture the stability properties of the chaotic dynamical systems with negligible errors, demonstrating the robustness of the framework.
\begin{table}[h]
\caption{Estimates of Lyapunov exponents and the Kaplan-Yorke dimension for different values of $L$.}\label{table:var_L_KS}
\begin{tabular*}{\textwidth}{@{\extracolsep\fill}lcccc}
\toprule%
& \multicolumn{2}{@{}c@{}}{$\lambda_1$} & \multicolumn{2}{@{}c@{}}{$D_{KY}$} \\\cmidrule{2-3}\cmidrule{4-5}%
$L$ & Reference & ESN & Reference & ESN \\
\midrule
$22$    & $0.046$ & $0.047$ & $6.007$  & $6.007$ \\
$44$    & $0.079$ & $0.080$ & $10.017$ & $10.017$ \\
$62.83$ & $0.085$ & $0.090$ & $15.018$ & $15.020$ \\
$76$    & $0.089$ & $0.100$ & $18.020$ & $18.030$ \\
\botrule
\end{tabular*}
\footnotetext{Comparison between the reference and CAE-ESN.}
\end{table}

\section{Turbulent flows: an application to 2D Navier-Stokes}\label{sec:kolmogorov-results}

Fluid turbulence can be viewed as a chaotic dynamical system with large numbers of degrees of freedom. As such, the two-dimensional Kolmogorov flow is a fundamental model to study two-dimensional turbulent flows. The system is governed by the incompressible Navier-Stokes equation

\begin{align}\label{eq:kolmogorov}
\partial_t \mathbf{u} &= -\mathbf{u}\cdot \nabla \mathbf{u} - \nabla p + v\Delta \mathbf{u} +\mathbf{f}, \\
\nabla \cdot \mathbf{u} &= 0. 
\end{align} 
where $\mathbf{u}: \Omega\times \mathbb{R}^{+} \to \mathbb{R}^2 $ represents the fluid velocity field, $p: \Omega\times \mathbb{R}^{+} \to \mathbb{R} $ is the pressure field, and $v = 1/ Re$, where $Re$ is the Reynolds number, denotes the viscosity. The forcing is defined by $\mathbf{f}(x) = \sin(k_fy)e_1 - \alpha \mathbf{u} $, with $k_f = (0, k_f)$ as the forcing wave number and $e_i $ represents the standard basis in $\mathbb{R}^2$. The second term is a velocity-dependent drag, controlled by the amplitude of $\alpha$ \citep{boffetta2012two, kochkov2021machine}. The flow is defined on the domain $\Omega = [0, 2\pi] \times [0, 2\pi] $ with periodic boundary conditions.
To solve Eq.~\eqref{eq:kolmogorov}, we employ a publicly available pseudospectral solver KolSol (\url{https://github.com/MagriLab/KolSol}) based on \citep{canuto1988spectral}.

After solving the equation in the Fourier domain, the solution is projected onto a $48 \times 48$ grid in the spatial domain. Through this discretization, the stability properties of the Kolmogorov flow can be studied from a dynamical system's perspective, as outlined in Section~\ref{sec:stability_properties}. For its transition from a quasiperiodic to a turbulent system, we refer to \citep{platt_kolmogorov_1991} and the covariant Lyapunov analysis can be found in \citep{inubushi2012covariant}. The bottleneck of stability analysis for turbulent flows is the calculation (and storing) of the Jacobian, and the subsequent solution of the variational equation. Evaluating the linearized right-hand side, especially when employing spectral solvers, is convoluted and inefficient and one typically employs Jacobian-free methods to calculate Lyapunov exponents \citep{dieci2002jacobian}. In the following, we consider two cases of the Kolmogorov flow: (i) a quasiperiodic solution of the system for $Re=30, \alpha=0.1$ \citep{racca2023predicting} and (ii) a turbulent solution of the system at $Re=34, \alpha=0.1$ \citep{kochkov2021machine}. 

\subsection{Quasiperiodic case}
The time-accurate and statistical prediction with the CAE-ESN for the quasiperiodic case has been studied in \citep{racca2023predicting}, and authors found good network performance for $N_{lat} \geq 9$. We generate snapshots of size $48\times 48$ with $\Delta t=0.1$ for $T=60000$, with $25\%$ considered for training, $12.5\%$ for validation and the remaining data for testing.

\begin{figure}[h]
    \centering
    \includegraphics[width=0.65\textwidth]{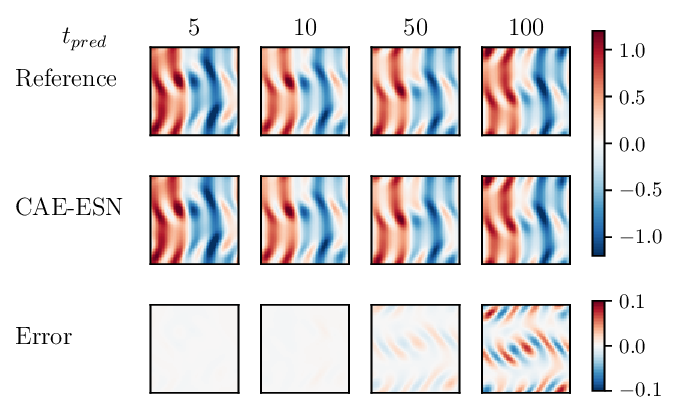} 
    \caption{Closed-loop prediction of the CAE-ESN compared to the reference data for the quasiperiodic Kolmogorov flow.}
    \label{fig:cae_esn_predre30}
\end{figure}

Following Section~\ref{subsec:cae}, we first train the autoencoder to generate the latent manifold with $N_{lat} = 18$, less than $1\%$ of the snapshot size. We then train the ESN on the latent observations $\mathbf{y}(t_i)$ spaced at $\Delta t = 1$. One closed-loop prediction over $100$ time units of the CAE-ESN on the test data is presented in Fig.~\ref{fig:cae_esn_predre30}. The closed-loop prediction correctly infers the quasiperiodic behaviour of the flow despite evolving only on $N_{lat} = 18$.

\begin{figure}[h]
    \centering
    \includegraphics[width=0.75\textwidth]{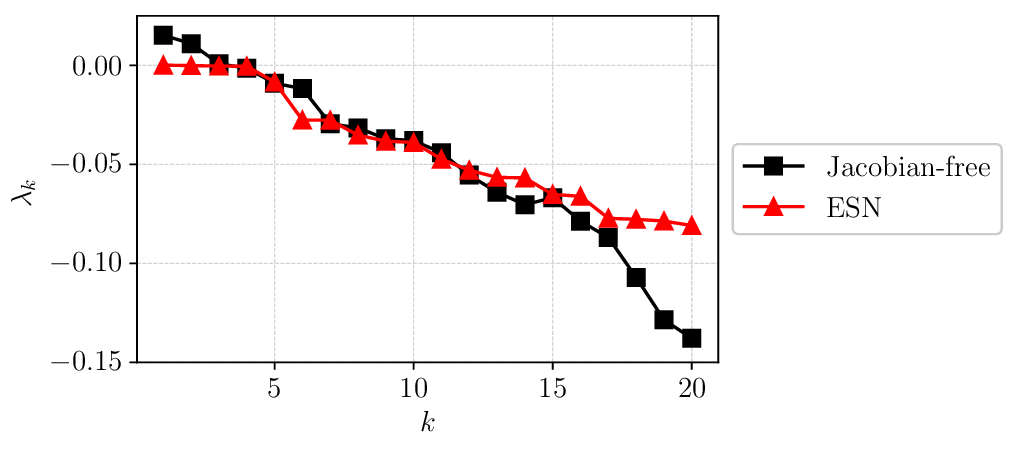} 
    \caption{Lyapunov spectrum of the CAE-ESN (red line) and the Jacobian-free method (black line) for a quasiperiodic solution of the Kolmogorov flow.}
    \label{fig:lyapunov_exponents_re30}
\end{figure}

To assess the quasiperiodic nature of the state evolution, we examine its stability by calculating the Lyapunov exponents using a Jacobian-free approach. In a quasiperiodic system, the largest LE should be $0$; however, using the Jacobian-free method, the largest LE converges to approximately $0.02$. This discrepancy arises because the Jacobian-free method provides an approximation of the spectrum. Figure.~\ref{fig:lyapunov_exponents_re30} shows a comparison of the Lyapunov spectrum of the CAE-ESN with the Jacobian-free method across the first $18$ exponents. The first four exponents from the CAE-ESN are on the order of $10^{-4}$, indicating that the network effectively captures the system's stable exponents in the latent space. Both methods show partial overlap in the negative spectrum, further validating the CAE-ESN's performance.

\subsection{Turbulent case}
For $Re=34, \alpha=0.1$, the Kolmogorov flow is turbulent and displays chaotic spatial and temporal behaviour. The vorticity data is generated on a $48\times48$ grid and sampled at $\Delta t =0.1$ for $T=20000$, with $25\%$ considered for training, $12.5\%$ for validation and the remaining data for testing. Using the Jacobian-free method, we estimate $\lambda_1 = 0.094$ and $D_{KY}=16.02$. 

\begin{figure}[h]
    \centering
    \includegraphics[width=0.65\textwidth]{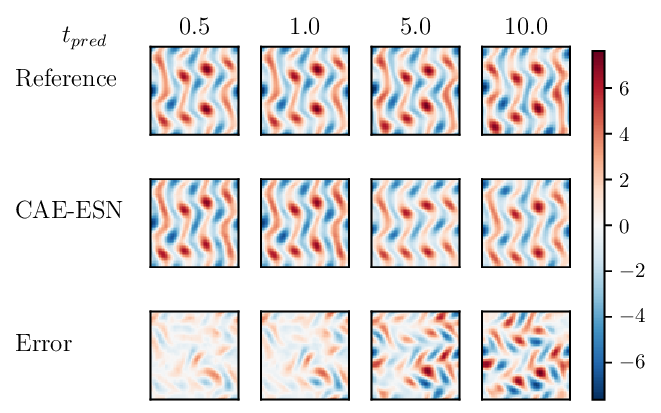} 
    \caption{Closed-loop prediction of the CAE-ESN compared to the reference for the turbulent case of the Kolmogorov flow.}
    \label{fig:cae_esn_predre34}
\end{figure}

After training the CAE for $N_{lat}=48$, the ESN is trained on the latent representations. In Fig.~\ref{fig:cae_esn_predre34}, we show the autonomous vorticity prediction of the CAE-ESN over $100$ time steps on unseen data. Compared to the reference solution, the CAE-ESN qualitatively captures analogous vortex structures. Due to chaos, errors accumulate in the closed-loop forecast; however, the accuracy of the ESN prediction is known to increase with the size of the reservoir \citep{racca2023predicting}. For this network architecture, we observe that after approximately $200$ time units, the autonomous forecast generated by the CAE-ESN becomes unstable, highlighting the inherent challenges of applying time series forecasting to turbulent systems \citep{linot2023stabilized,racca2023predicting}. 

\begin{figure}[h]
    \centering
    \includegraphics[width=0.75\textwidth]{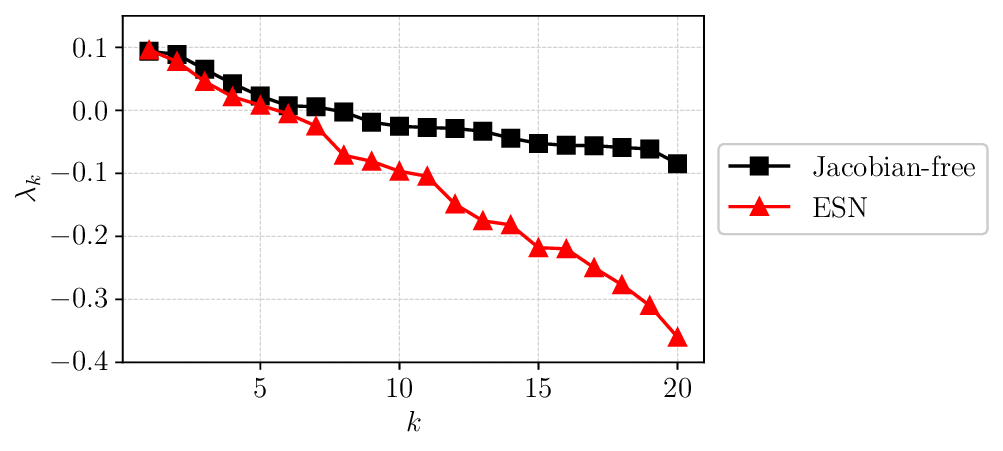} 
    \caption{Lyapunov spectrum of the CAE-ESN (red line) and the Jacobian-free method (black line) for the turbulent Kolmogorov flow.}
    \label{fig:lyapunov_exponents_re34}
\end{figure}

Following \citep{racca2023predicting}, we discard the remaining part of the closed-loop prediction if the prediction exceeds the range of the training set. The comparison of the resulting Lyapunov exponents with those obtained via the Jacobian-free method, as shown in Fig.~\ref{fig:lyapunov_exponents_re34}, reveals good agreement in the dominant exponent, with $\lambda_1^{CAE-ESN}=0.096$ compared to $\lambda_1=0.094$ of the reference. The positive spectrum aligns well, whilst discrepancies are observed in the negative exponents, which correspond to the system's stable directions. This is because negative exponents are generally more challenging to extract numerically, and the Jacobian-free reference provided above merely serves as an estimate rather than an exact solution. In general, the reliable estimation of negative Lyapunov exponents is often not critical for diagnosing chaotic systems, as the dynamics are primarily characterized by the positive Lyapunov exponents\citep{yang2011robust}. While both methods are challenged in accurately capturing the negative spectrum of turbulent flows, this highlights the potential of the CAE-ESN effectively capturing the dominant features of the dynamical system in a reduced manifold.

\section{Conclusion} \label{sec:conclusion}
In this work, we employ a convolutional autoencoder (CAE) combined with an echo state network (ESN) for the stability analysis of spatial-temporal chaos. The encoder reduces the full state of the dynamical system onto a low-dimensional latent manifold, on which the ESN propagates the temporal dynamics. The ESN prediction can then be decoded to obtain the full physical state. First, we demonstrate that the propagation on the latent manifold leads to good agreement in the short-term forecast. Second, we perform stability analysis based on the Jacobian of the ESN, hence calculating the tangent space on the latent manifold. We show that key stability properties of the Kuramoto-Sivashinsky equation are captured in the low-dimensional representation by calculating Lyapunov exponents, the Kaplan-Yorke dimension, and the covariant Lyapunov exponents. The inferred Kaplan-Yorke dimension of the latent manifold has excellent agreement with the reference dimension of the attractor. The learned properties are robust across different parameter setups and correspond to the properties retrieved from the Jacobian of the numerical solution. 
Finally, we extend the CAE-ESN to solutions of the Navier-Stokes equation. We compare the Lyapunov spectrum of the CAE-ESN with Jacobian-free approximation methods on the Kolmogorov flow for two stability cases (i) a quasiperiodic case and (ii) a turbulent case, resulting in a good approximation of the Lyapunov spectra. 
This work analyses the stability properties of a low-dimensional manifold generated by the CAE-ESN. Firstly, the CAE-ESN framework provides a tool to accurately infer stability properties from high-dimensional data when the governing equations are not known. Furthermore, by showing that the network's stability properties are compatible with the physical system, the CAE-ESN proves to be an effective tool for modelling chaotic dynamics. This work opens new opportunities for analysing the stability properties of high-dimensional chaotic systems through data-driven reduced-order models.

\bmhead{Acknowledgements}
The authors thank Defne Ege Ozan for insightful discussions regarding the ESN and Alberto Racca for valuable suggestions regarding the data-driven modelling of the Kolmogorov flow. This research has received financial support from the EPSRC Grant No. EP/W026686/1 and from the ERC Starting Grant No. PhyCo 949388.

\backmatter

\begin{appendices}
\section{Appendix}\label{sec:appendix}
\subsection{Architecture of Autoencoders}\label{sec:architecture_hyperparameter}
In this section, we elaborate on the different autoencoder architectures. All autoencoders were implemented in PyTorch\footnote{\url{https://pytorch.org/}} and trained on a single NVIDIA RTX 8000. A tutorial for the implementation is available on GitHub\footnote{\url{https://github.com/MagriLab/LatentStability}}.
\begin{table}[h]
\begin{tabular}{ p{3cm}  p{4cm} } 
  \hline
  Encoder layer & Dense \\ 
  \hline 
    Decoder layer & Dense \\ 
  \hline
  $N_{lat}$ & $6, 8, 12, 16, 20, 24, 28, 32$\\ 
  \hline
  Activation function & $tanh$ \\ 
  \hline
\end{tabular}\caption{Hyperparameters of the CAE}\label{table:cae_dense} 
\end{table}

\begin{table}[h]
\begin{tabular}{ p{3cm}  p{4cm}  } 
  \hline
  Encoder layers &   \begin{tabular}{@{}l@{}}  4 Conv1D (with tanh); \\ 1 Dense  \end{tabular} 
   \\ 
  \hline 
   Decoder layers &  \begin{tabular}{@{}l@{}}    1 Dense; \\4 Conv1D (with tanh) \end{tabular} \\ 
  \hline 
  \hline
    Channels &  $2 \to 4 \to 8  \to16 $ \\ 
  \hline
    Filters &  $8 \to 8\to 3  \to 3$ \\ 
  \hline
     Strides &  $1 \to 2 \to 2  \to 2$ \\ 
  \hline
\end{tabular}\caption{Parameters of the CAE for input dimension $N_x=128$}\label{table:cae_128} 
\end{table}

\begin{table}[h]
\begin{tabular}{ p{3cm}  p{4cm} } 
  \hline
  Encoder layers &  \begin{tabular}{@{}l@{}} 7 Conv1D (with tanh); \\ 1 Dense  \end{tabular}   \\ 
  \hline 
   Decoder layers & \begin{tabular}{@{}l@{}}    1 Dense; \\7 Conv1D (with tanh) \end{tabular}\\ 
  \hline 
  \hline
    Channels &  $2 \to 4 \to 8  \to16  \to 32 \to 64$ \\ 
  \hline
    Filters &  $8 \to 8 \to 5  \to 5  \to 3  \to 3$ \\ 
  \hline
     Strides &  $1 \to 2 \to 2  \to 3  \to 3  \to 3$ \\ 
  \hline

\end{tabular}\caption{Parameters of the CAE for input dimension $N_x=512$}\label{table:cae_512} 
\end{table}

\subsection{Training of Echo State Networks}\label{sec:appendix_esn}
To train the ESN, we follow \citep{racca2021robust}: First, we fix the reservoir size and the parameter range of the parameter of the ESN, see Table ~\ref{table:esn_parameters}. After an initial parameter grid search, a Bayesian optimization is employed to find hyperparameters which minimize the validation loss. 
\begin{table}[h]
\begin{tabular}{  p{4cm}  p{3cm} } 
  \hline
  Spectral radius $\frenchspacing$ &   $[0.1, 1.25]$   \\ 
  \hline 
   Input scaling $\sigma_{in}$ &  $[0.01, 10.0]$\\ 
  \hline 
   Tikhonov regularization  $\beta$&  $ [10^{-3}, 10^{-11}]$\\
  \hline
  \hline
    Connectivity $d$ & $10$ \\ 
  \hline
    Input bias $b_{in}$ &  $1$ \\ 
  \hline
    Output bias $b_{out}$ &  $1$ \\ 
  \hline

\end{tabular}\caption{Parameters of the ESN }\label{table:esn_parameters} 
\end{table}

\subsection{Parameter study}

\subsubsection{$L=22$}

The case $L=22$ is the most commonly considered case of the KS equation \citep{gupta2023mori, vlachas2022multiscale, linot2020deep, bucci2019control}. The system is chaotic in a Lyapunov sense and converges toward a bimodal steady attractor \citep{papageorgiou1991route}. The manifold's dimension $d_{M}$ is estimated to be of $d_{M}=8$ \citep{linot2020deep} whilst the attractor's dimension is estimated at $d_{KY}=6.001$. The system is solved with $N_x=256, \Delta t= 0.2$ following \citep{Kassam_2005_fourth_order} and exhibits a largest Lyapunov exponent of $\lambda_1= 0.046$. The system has two positive Lyapunov exponents and two zero Lyapunov exponents with the remaining exponents being negative. 
We train CAEs with varying latent dimensions of $N_{lat} \in [ 6, 8, 10, 12]$ and perform hyperparameter optimisation of the ESNs with $N_r = 5000$ in the respective latent spaces. The details on the parameters can be found in Table~\ref{table:cae_512} and Table~\ref{table:esn_parameters}. 

\begin{figure}[h]
    \centering
    \includegraphics[width=0.75\textwidth]{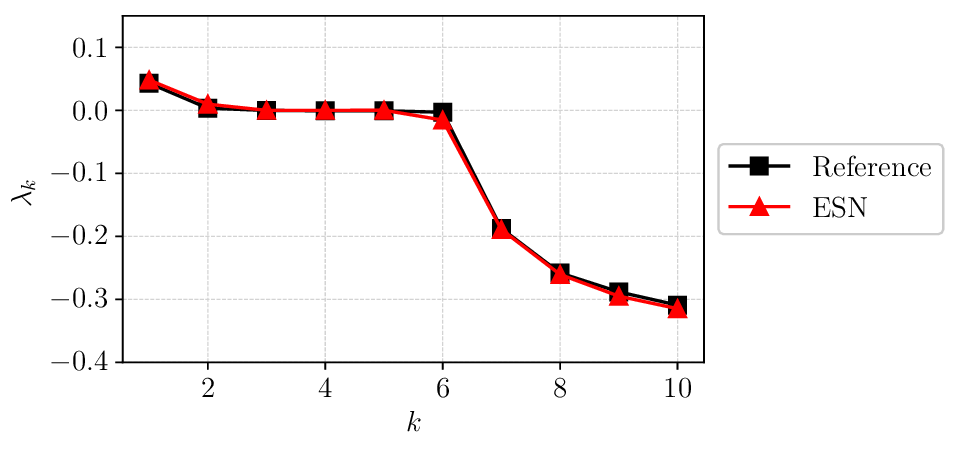} 
    \caption{Comparison between the Lyapunov exponents of the reference data at $L=22$ (black line) and those calculated from the echo state network with $N_{lat} = 8$ (red line).}
    \label{fig:LEs_L22_ESN}
\end{figure}

In Fig.~\ref{fig:LEs_L22_ESN}, we present the Lyapunov exponents of the ESN in the latent space of dimension $8$ compared to the reference data. We comment here that despite extensive hyperparameter testing, the ESN was not able to recover the Lyapunov exponents and the Kaplan-Yorke dimension when selecting a latent dimension smaller than the inertial manifold's dimension. 

\begin{figure}[h]
    \centering
    \includegraphics[width=0.7\textwidth]{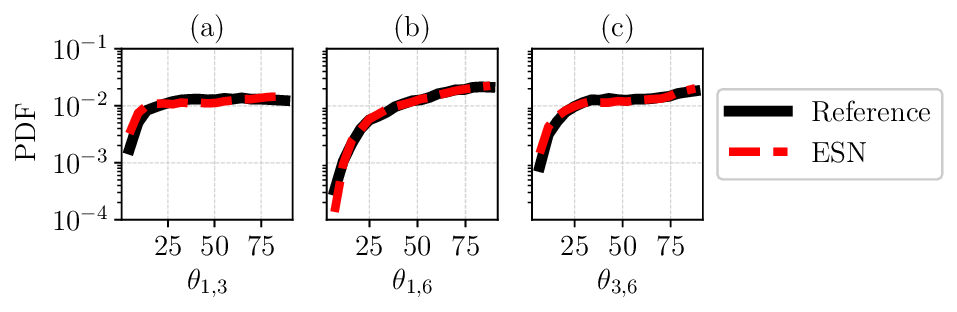} 
    \caption{Statistics of the angles between selected pairs of CLVs from the reference data at $L=22$  (black line) and the echo state network with $N_{lat} = 8$ (red line). }
    \label{fig:CLVs_L22_ESN}
\end{figure}

When selecting a latent dimension of $8$ and larger, performing the stability analysis in the latent space also leads to good agreement in the covariant Lyapunov vector angles, as demonstrated in Fig.~\ref{fig:CLVs_L22_ESN}. We demonstrated robustness towards different ESN realizations for this case in \citep{ozalp2024neurips}.

\subsubsection{$L=44$}
\begin{figure}[h]
    \centering
    \includegraphics[width=0.7\textwidth]{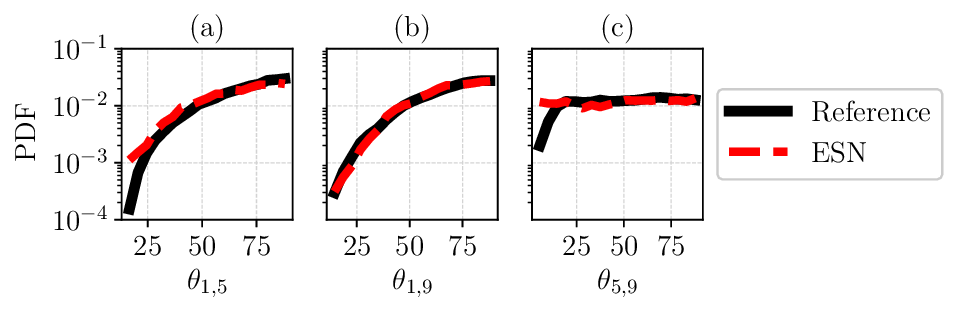} 
    \caption{Statistics of the angles between selected pairs of CLVs from the reference data at $L=44$ (black line) and the echo state network with $N_{lat} = 18$ (red line). }
    \label{fig:CLVs_L44_ESN}
\end{figure}

For $L=44$, we solve the system with $N_x=256, \Delta t= 0.1$ and obtain $\lambda_1 = 0.079, D_{KY}=10.01$. Linot and Graham report the corresponding manifold's dimension at $d_{M} = 18$ \citep{linot2022data}. The CAE implementation follows Table~\ref{table:cae_512} for $N_{lat} \in [12, 16, 18, 20, 24, 28, 32]$ and the ESN is trained according to Table~\ref{table:esn_parameters} with $N_{r}=5000$. The Lyapunov exponents of the ESN applied to a latent space of size $18$ are presented in Fig.~\ref{fig:LEs_L22_ESN}, as well as the angle statistics of the CLVs in Fig.~\ref{fig:CLVs_L44_ESN}.
\begin{figure}[h]
    \centering
    \includegraphics[width=0.75\textwidth]{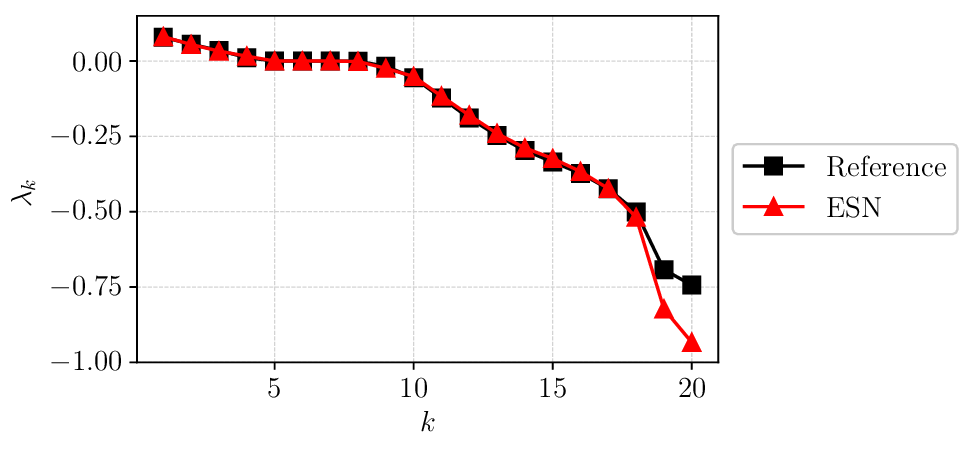} 
    \caption{Comparison between the Lyapunov exponents of the reference data at $L=44$ (black line) and those calculated from the echo state network with $N_{lat} = 18$ (red line). }
    \label{fig:LEs_L44_ESN}
\end{figure}

\subsubsection{$L=76$}
\begin{figure}[h]
    \centering
    \includegraphics[width=0.75\textwidth]{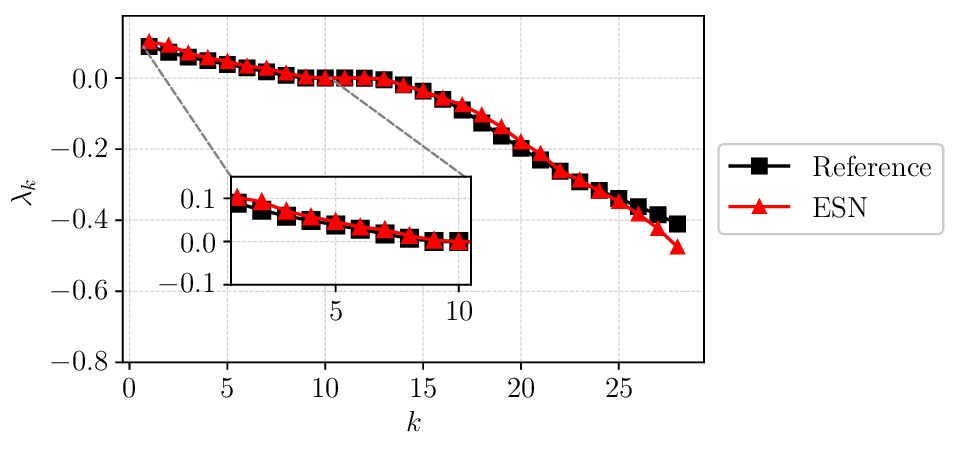} 
    \caption{Comparison between the Lyapunov exponents of the reference data at $L=76$ (black line) and those calculated from the echo state network in the latent space (red line). }
    \label{fig:LEs_L76_ESN}
\end{figure}

\begin{figure}[h]
    \centering
    \includegraphics[width=0.7\textwidth]{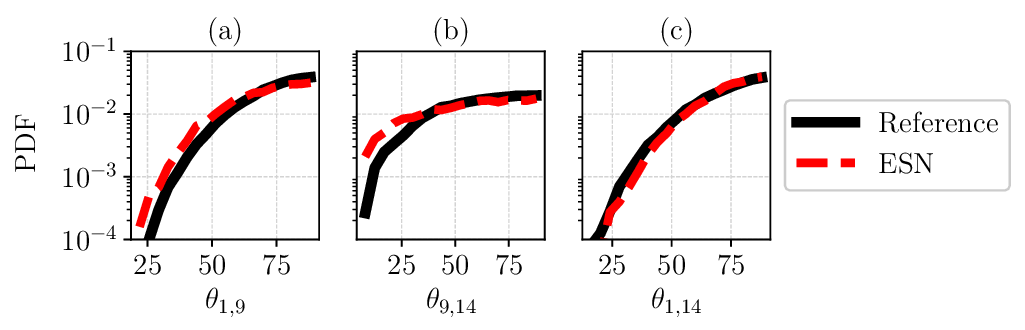} 
    \caption{Statistics of the angles between selected pairs of CLVs from the reference data at $L=76$ (black line) and the echo state network in the latent space of dimension $28$ (red line). }
    \label{fig:CLVs_L76_ESN}
\end{figure}

The KS equation with $L=76$ is solved with $N_x=128, \Delta t= 0.2$, resulting in $\lambda_1 = 0.089, D_{KY}=18.02$. The implementation follows Table~\ref{table:cae_128} for $N_{lat} \in [24, 28, 32, 36, 40]$ and Table~\ref{table:esn_parameters} with $N_r = 15000$.
The Lyapunov exponents in Fig.~\ref{fig:LEs_L76_ESN} and angle statistics of the CLVs in Fig.~\ref{fig:CLVs_L76_ESN} are obtained in a latent space with $N_{lat} = 28$.

\end{appendices}


\bibliography{sn-bibliography}

\end{document}